\documentclass[ reprint, superscriptaddress,
bibnotes, amsmath,amssymb, aps, prl
pra, prb, rmp, prstab, prstper, floatfix,
]{revtex4-2}

\usepackage{graphicx}
\usepackage{dcolumn}
\usepackage{bm}
\usepackage[super]{nth} 
\usepackage{dsfont} 

\usepackage{orcidlink} 
\usepackage{mathtools} 
\usepackage{cleveref} 
\usepackage{xcolor} 

\newcommand{\average}[1]{\left\langle#1\right\rangle}
\newcommand{\ket}[1]{\left\vert#1\right\rangle}
\newcommand{\bra}[1]{\left\langle#1\right\vert}
\newcommand{\comm}[2]{\left[#1,#2\right]}

\newcommand{\pTrace}[2]{\underset{#1}{Tr}\left[#2\right]}

\newcommand{\oper}[2]{\left\vert#1\rangle\!\langle#2\right\vert}
\newcommand\redsout{\bgroup\markoverwith{\textcolor{red}{\rule[0.5ex]{2pt}{0.4pt}}}\ULon}

\definecolor{ppblue}{RGB}{46,117,182}
\definecolor{ppred}{RGB}{197, 90, 17}

\begin{document}
\preprint{}
\title{Memory effects in repeated uses of quantum channels}
\author{Hayden Zammit\,\orcidlink{0009-0003-9917-929X}}
\affiliation{Department of Physics, University of Malta, Msida MSD 2080, Malta}

\author{Roberto Benjamin Salazar Vargas\,\orcidlink{0000-0003-1737-1433}}
\affiliation{Department of Physics, University of Malta, Msida MSD 2080, Malta}

\author{Gianluca Valentino\,\orcidlink{0000-0003-3864-7785}}
\affiliation{Department of Communications and Computer Engineering, University of Malta, Msida MSD 2080, Malta}

\author{Johann A. Briffa\,\orcidlink{0000-0003-3532-9474}}
\affiliation{Department of Communications and Computer Engineering, University of Malta, Msida MSD 2080, Malta}

\author{Tony J. G. Apollaro\,\orcidlink{0000-0002-9324-9336}}
\affiliation{Department of Physics, University of Malta, Msida MSD 2080, Malta}

\date{\today}

\begin{abstract}
    Quantum Information Processing (QIP) tasks can be efficiently formulated in
    terms of quantum dynamical maps, whose formalism is able to provide the
    appropriate mathematical representation of the evolution of open quantum
    systems. A key QIP task is quantum state transfer (QST) aimed at sharing
    quantum information between distant nodes of a quantum network, enabling,
    e.g.\ quantum key distribution and distributed quantum computing. QST has
    primarily been addressed insofar by resetting the quantum channel after each
    use, thus giving rise to memoryless channels. Here we consider the case
    where the quantum channel is continuously used, without implementing time- and
    resource-consuming resetting operations. We derive a general, analytical
    expression for the $n\nthscript{th}$-use average QST fidelity for $U(1)$-symmetric
    channels and apply our formalism to a perfect QST channel in the presence of
    imperfect readout timing. We show that even relatively small readout timing
    errors give rise to memory effects which have a highly detrimental impact on
    subsequent QST tasks.
\end{abstract}

\maketitle

\section{Introduction}
Quantum Information Processing (QIP) generally
requires a precise sequence of perfectly timed and controlled operations.
    Control of qubit interactions for
    quantum gate operations~\cite{Koch2022}  and  accurate synchronization of
    input–output procedures such as state preparation and
    readout~\cite{robledoHighfidelityProjectiveReadout2011a} are key for the successful implementation of a
    QIP protocol.
    Despite remarkable
    progress in time-controlled operations, both experimental imperfections and
    fundamental limits on timekeeping precision~\cite{meierFundamentalAccuracyresolutionTradeoff2023}
    prevent perfect temporal control. Consequently, even minor timing mismatches can
    induce significant errors in the performance of QIP tasks, motivating extensive
    research on the impact of timing
    errors~\cite{xuerebImpactImperfectTimekeeping2023} and on the energetic cost of
    achieving high temporal accuracy~\cite{meierPrecisionNotLimited2025b}. In this
    work, we combine these two perspectives by investigating how many times a quantum channel can be sequentially
    used to perform a fundamental QIP task, the
    quantum state transfer (QST) of a single qubit, before an
    energetic cost needs to be spent in order to reset it to its initial
    state because of detrimental memory effects~\cite{navascuesResettingUncontrolledQuantum2018,bassmanoftelieDynamicCoolingContemporary2024}.

Although single-use QST is well studied~\cite{Bose2003,
        Nikolopoulos2014, 10.1063/1.4978327, apollaroEntangledStatesAre2022},
    repeated-use scenarios remain poorly understood due to memory effects generated across uses. Understanding how such effects alters the
    channel's dynamics and information capacity is vital for scalable quantum
    networks, making memory-full channels a central focus of current
    research~\cite{carusoQuantumChannelsMemory2014b}.
    Their proper characterization is essential for enhancing signal throughput
    in optical fibers~\cite{10663983} and mitigating cross-talk noise in
    solid-state quantum processors~\cite{sarovarDetectingCrosstalkErrors2020}.
    While the impact of memory effects has been widely explored in optical
    systems \cite{benentiEnhancementTransmissionRates2009,
        carusoTeleportationInducedCorrelatedQuantum2010,
        meleRestoringQuantumCommunication2022a}, they remain largely uncharted in
    solid-state based channels. Notably, Ref.~\cite{bayatMemoryEffectsSpinchain2008a},
    analyzed a four-site Heisenberg spin-$\frac{1}{2}$ chain that under two
    successive uses demonstrated that memory effects can enhance both classical
    and quantum capacities compared to the memoryless case. Similarly,
    Ref.~\cite{PhysRevA.79.012311}, proposed a QST protocol that remains
    reliable in the limit of many uses, provided that the information from the
    sender qubit is distributed among multiple receivers. These approaches,
    together with the present work, belong to the class of perfect memory
    channels ~\cite{PhysRevA.69.012306,Giovannetti_2005}.

The paper is organized as follows: in Sec.~\ref{sec_model} we develop a general analytical framework for the
    dynamics of quantum channels used repeatedly without resetting; in Sec.~\ref{sec_results}  we apply
    our formalism to  spin-$\frac{1}{2}$ quantum channels with  $U(1)$ symmetry,
    obtaining an exact expression for the $n\nthscript{th}$-use average fidelity; in Sec.~\ref{sec_PST} we focus on a perfect state
    transfer (PST) spin-$\frac{1}{2}$ model determining the impact of memory effects on the QST quality and providing an upper bound on the quantum channel capacity; finally, in Sec.~\ref{sec_concl} we draw our conclusions.

\section{Model}\label{sec_model}
\begin{figure}
    \centering
    \includegraphics[width=\linewidth]{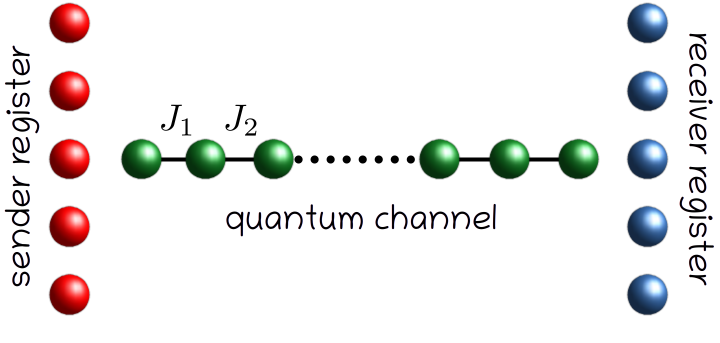}
    \caption{A quantum channel (green spheres) for the transfer of quantum information between two quantum registers: a sender register (red spheres) and a receiver register (blue spheres). }
    \label{fig_model}
\end{figure}

We consider the following protocol for the transfer of the information from a sender register to a receiver register as illustrated in Fig.~\ref{fig_model}:
\begin{enumerate}
    \item The channel is initialized in $\ket{\Psi}_{C_1}=\ket{000\dots000}$. At
          time $t_0=0$, the \nth{1} qubit of the sender register, in an arbitrary
          state
          $\ket{\Psi}_{S_1}=\cos\frac{\theta_1}{2}\ket{0}+\sin\frac{\theta_1}{2}e^{i
              \phi_1}\ket{1}$, and of the receiver register, in the state $\ket{0}_{R_1}$,
          are attached to the channel;
    \item The system $S+C+R$ evolves for a time $t_1$, when the
          \nth{1} sender and receiver qubits are
          substituted with the \nth{2} set,  respectively in the states
          $\ket{\Psi}_{S_2}=\cos\frac{\theta_2}{2}\ket{0}+\sin\frac{\theta_2}{2}e^{i
              \phi_2}\ket{1}$ and $\ket{0}_{R_2}$. The whole system is then let to
          evolve for a time $t_2$.
    \item This procedure is repeated until the whole sender register is
          transferred to the receiver register.
\end{enumerate}

Our figure of merit is the average $n\nthscript{th}$-use QST fidelity
without resetting the quantum channel to $\ket{\Psi}_{C_1}$. We define
the $n\nthscript{th}$-use average fidelity:
\begin{align}\label{eq_avFn}
    \average{F_n\left(t_n;t_1\right)}=\int d\psi\bra{\Psi}_{S_{n}}\hat{\rho}_R^{(n)}\left(t_n;t_1\right)\ket{\Psi}_{S_{n}}~,
\end{align}
where the state of the $n\nthscript{th}$-use receiver is given by
\begin{align}
    \hat{\rho}_R^{(n)}\left(t_n;t_1\right) & =\pTrace{S,C}{\hat{\rho}_{SRC}\left(t_n;t_1\right)} \\
                                       & =\pTrace{S,C}{\hat{U}(t_n)\hat{\rho}_{S}^{(n)}\hat{\rho}_{C}^{(n-1)}\left(t_{n-1};t_1\right)\hat{\rho}_{R}^{(n)}\hat{U}^{\dagger}(t_n)}\nonumber,
\end{align}
and the state of the channel is given by the output of the previous use
\begin{align}
     & \hat{\rho}_C^{(n-1)}\left(t_{n-1};t_1\right)=\pTrace{S,R}{\hat{\rho}_{SRC}\left(t_{n-1};t_1\right)}                                                                           \\
     & =\int d\psi~\pTrace{S,R}{\hat{U}(t_{n-1})\hat{\rho}_{S}^{(n-1)}\hat{\rho}_{C}^{(n-2)}\left(t_{n-2};t_1\right)\hat{\rho}_{R}^{(n-1)}\hat{U}^{\dagger}(t_{n-1})}\nonumber
\end{align}
where $\left(t_{n-1};t_1\right)\equiv\left(t_{n-1},t_{n-2},\dots, t_2, t_1\right)$. The integration is done over the Bloch sphere with respect to the normalized Haar
measure, which accounts for the standard scenario where the sender states of previous channel uses are unknown.

\section{Results}\label{sec_results}
Our main result is the following exact expression for the
$n\nthscript{th}$-use average fidelity 
in \cref{eq_avFn} for $U(1)$-symmetric
quantum channels, i.e., when $\comm{\hat{H}}{\hat{M}_z}=0$, when $\hat{H}$ can be mapped to quadratic fermion models:
\begin{align}
    \label{eq_fid_gen}
    \average{F_n\left(t_n;t_1\right)}=\frac{1}{2}+\frac{\left|f_1^N\left(t_n\right)\right| A_{n-1}\left(t_{n-1};t_1\right)}{3}+\frac{\left|f_1^N\left(t_n\right)\right|^2}{6}~,
\end{align}
where $f_i^j\left(t_k\right)=\bra{j}\hat{U}(t_k)\ket{i}$ are
    transition amplitudes and $A_{n-1}\left(t_{n-1};t_1\right)$ accounts
for the memory effects acquired during each use. Eq.~\ref{eq_fid_gen} is derived from the average $n\nthscript{th}$-use fidelity of general
$U(1)$-symmetric quantum channels given in~\cref{eq.general-avg} in the Appendix. Notably, the
structure of the $n\nthscript{th}$-use average fidelity is independent of the number of
uses $n$ and equals the structure of the $\nth{1}$-use expression in Ref.~\cite{Bose2003}.

\begin{figure}
    \centering
    \includegraphics[width=\linewidth]{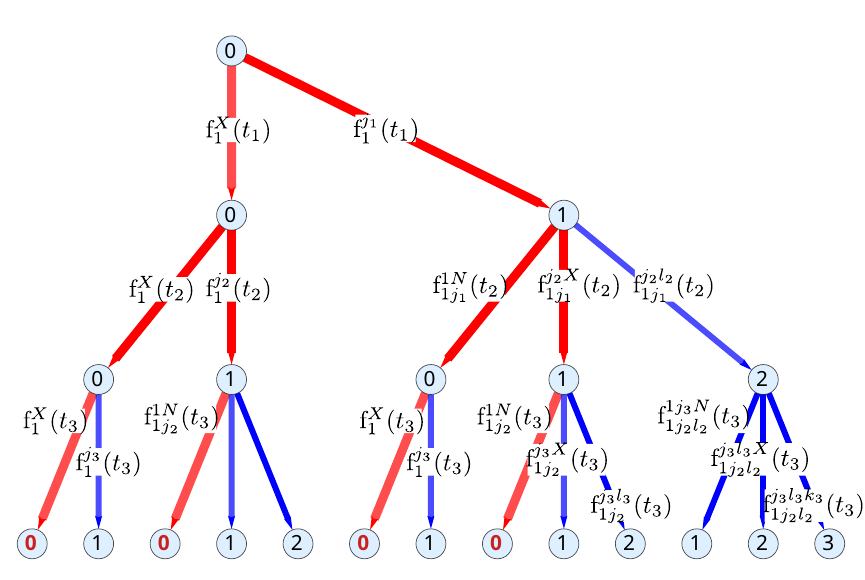}
    \caption{Mixed-degree rooted tree showing the transition amplitudes entering
        the term $A_{n-1}\left(t_{n-1};t_1\right)$ in \cref{eq_fid_gen}. The
        relevant transition amplitudes are along the Motzkin paths marked in red.
        Each vertex (blue dot) shows the number of excitations inside the channel
        after the readout procedure. The edges represent transition amplitudes
        between the states with the number of excitations between connected vertices. }
    \label{fig:graph}
\end{figure}

The memory effects, encoded into $A_{n-1}\left(t_{n-1};t_1\right)$, are
determined by summing over all the paths in the mixed-degree rooted tree graph
in Fig.~\ref{fig:graph} that end in 0 and multiplying all the transition
amplitudes at times $\left(t_{n-1};t_1\right)$. The graph is constructed by
labeling each node with integers $n\geq 0$ and applying the generation rule
\begin{align}
    G(n)=\begin{cases}
             \{0,1\}~       & n=0 \\
             \{n-1,n,n+1\}~ & n>0
         \end{cases}~
\end{align}
at each readout time $t_i$. Using the graph in Fig.~\ref{fig:graph}, it is
straightforward to determine the $n\nthscript{th}$-use average fidelity. As an
illustrative case, we derive the memory effect for the $\nth{4}$-use. The term
$A_{3}\left(t_3,t_2,t_1\right)$ in \cref{eq_fid_gen} is obtained as the sum of
the four Motzkin ~\cite{DonagheyShapiro1977MotzkinNumber} paths:
\begin{itemize}
    \item $0000 \rightarrow \displaystyle
              \prod_{i=1}^3\sum_{X_i\in\{1,N\}}\left|f_1^{X_i}(t_i)\right|^2$
    \item $0100 \rightarrow \displaystyle\left|\sum_{j_1\in
                  C}f_1^{j_1}(t_1)f_{1j_1}^{1N}(t_2)\right|^2\sum_{X\in\{1,N\}}\left|f_1^{X}(t_3)\right|^2$
    \item $0010 \rightarrow \displaystyle
              \sum_{X\in\{1,N\}}\left|f_1^X(t_1)\right|^2\left|\sum_{j_2\in
                  C}f_1^{j_2}(t_2)f_{1j_2}^{1N}(t_3)\right|^2$
    \item $0110 \rightarrow \displaystyle
              \sum_{X\in\{1,N\}}\left|\sum_{j_1,j_2\in
                  C}f_1^{j_1}(t_1)f_{1j_1}^{j_2X}(t_2)f_{1j_2}^{1N}(t_3)\right|^2$
\end{itemize}
The quadratic property of the Hamiltonian can then be utilized to further reduce
\cref{eq_fid_gen} in terms of only single-particle transition
amplitudes~\cite{apollaroQuantumTransferInteracting2022} using Slater
determinants. For instance, the product of single- and two-particle transition
amplitudes appearing in the $0100$ Motzkin path transforms as follows into
products of single-particle amplitudes
\begin{align}
    \label{eq_single_particle}
    \sum_{j_1\in C}f_1^{j_1}(t_1)f_{1j_1}^{1N}(t_2)=\sum_{j_1\in C}f_1^{j_1}(t_1)\begin{vmatrix}
    f_{1}^{1}(t_2)   & f_{1}^{N}(t_2)   \\
    f_{j_1}^{1}(t_2) & f_{j_1}^{N}(t_2)
    \end{vmatrix}~
\end{align}
Furthermore, exploiting the completeness relation
$\sum_{i=1}^{N}\oper{i}{i}=\mathds{1}$, the sum over the channel's spins can be
eliminated. The term arising by expanding the determinant in
\cref{eq_single_particle} can be expressed in terms of transition
amplitudes involving only the sender and the receiver,
\begin{align}
    \label{eq_single_particle1}
     & \sum_{j_1\in C}f_{j_1}^{N}(t_2)f_1^{j_1}(t_1)\equiv \sum_{j_1\in C}\bra{N}\hat{U}(t_2)\oper{j_1}{j_1}\hat{U}(t_1)\ket{1}\nonumber \\
     & =\bra{N}\hat{U}(t_2)\left(\mathds{1}-\oper{1}{1}-\oper{N}{N}\right)\hat{U}(t_1)\ket{1}                                            \\
     & =f_1^N(t_1+t_2)-f_1^1(t_1)f_1^N(t_2)-f_1^N(t_1)f_N^N(t_2)\nonumber \\ 
 &\coloneq B_1^N(t_2,t_1)=B_2\nonumber
\end{align}
Notably, the procedure outlined in \cref{eq_single_particle1} entails that the
$n\nthscript{th}$-use average fidelity can be determined, in spin-$\frac{1}{2}$
Hamiltonians that map to free fermion models~\cite{Lieb1961}, just by means of
the transition amplitudes involving the sender and the receiver. This finding
constitutes our second main result.

To showcase the usability and the convenience of our approach for repeated
uses of a quantum channel for QST, let us report a few $n\nthscript{th}$-use scenarios.
Clearly, our result coincides with the first-use average fidelity
$\average{F_1(t)}$ given in Ref.~\cite{Bose2003}
\begin{align}
    \average{F_1(t_1)}=\frac{1}{2}+\frac{\left|f_1^N(t_1)\right|}{3}+\frac{\left|f_1^N(t_1)\right|^2}{6}~,
\end{align}
as there is no Motzkin path at the top of the graph in Fig.~\ref{fig:graph}
where $A_0(t_0)=1$. The $\nth{2}$-use average fidelity reads
\begin{align}
    \label{eq_Favg2}
     & \average{F_{2}(t_2, t_1)}=\frac{1}{2}+\frac{\left|f_1^N(t_2)\right|A_1(t_1)}{3}+\frac{\left|f_1^N(t_2)\right|^2}{6}~,
\end{align}
with $A_1(t_1)=\left|f_1^1(t_1)\right|^2+\left|f_1^N(t_1)\right|^2$, and the
$\nth{3}$-use average fidelity
\begin{align}
    \label{eq_Favg3}
     & \average{F_3(t_3,t_2, t_1)}=\frac{1}{2}+\frac{1}{3}\left|f_1^N(t_3)\right|A_2(t_2,t_1)+\frac{\left|f_1^N(t_3)\right|^2}{6}~,
\end{align}
where
\begin{align}
     & A_2(t_2,t_1)=\left(\left|f_1^1(t_1)\right|^2+\left|f_1^N(t_1)\right|^2\right)\left(\left|f_1^1(t_2)\right|^2+\left|f_1^N(t_2)\right|^2\right)\nonumber \\
     & +\left|f_1^N(t_2)\left(f_1^1(t_1+t_2)-f_1^N(t_1)f_N^1(t_2)\right)\right.\nonumber                                                                      \\
     & \left.-f_1^1(t_2)\left(f_1^N(t_1+t_2)-f_1^N(t_1)f_N^N(t_2)\right)\right|^2~.
\end{align}
Higher $n\nthscript{th}$-uses can be similarly expressed and it is evident that the
memory effects arising from previous uses, embodied in
$A_{n-1}(t_{n-1};t_1)$ can only be detrimental on the average fidelity for equal readout timing protocols, i.e., $t_{n}=t_{n-1}=\dots=t_1$, as $0\leq
    A_{n-1} \leq A_{n-2}\dots\leq A_1\leq 1$. The left equality holds when all previous
uses have resulted in the excitation being trapped with unit probability inside
the channel at previous reading times $\left\{t_{i}\right\}$, whereas the right
equality holds when all the excitations have been extracted from the chain
successfully at previous reading times. Indeed, the latter scenario corresponds to an effective resetting of the quantum channel.
\section{PST chain}\label{sec_PST}
Here we consider spin-$\frac{1}{2}$ chains able to perform PST. Several instances have been proposed based on engineered
nearest-neighbor couplings~\cite{Christandl2004,vinetHowConstructSpin2012}
and next-nearest neighbor couplings~\cite{christandlAnalyticNexttonearestneighborModels2017} in the $XX$
Hamiltonian. We focus on the  coupling scheme of Ref.~\cite{Christandl2004} 
where the PST time is $\tau=\frac{\pi}{2}$. The Hamiltonian reads
\begin{align}
    \label{eq_Chris}
    \hat{H}=\sum_{i=1}^{N-1}J_i\left(\hat{\sigma}_i^x \hat{\sigma}_{i+1}^x+\hat{\sigma}_i^y \hat{\sigma}_{i+1}^y \right)~,
\end{align}
where $J_i=\sqrt{i(N-i)}$. Experimental realizations of PST with the model in
\cref{eq_Chris} have been realized in superconducting qubits~\cite{Li2018a} and
optical waveguides ~\cite{chapmanExperimentalPerfectState2016a}. The Hamiltonian
in \cref{eq_Chris} maps to a free fermion model and, hence, only single-particle
transition amplitudes involving the edge sites enter the $n\nthscript{th}$-use average
fidelity. Furthermore, these allow an explicit representation in terms of
reduced Wigner $d$-function~\cite{christandlPerfectTransferArbitrary2005},
yielding $f_1^1(t)=f_{N}^{N}(t)=\left(\cos t\right)^{N-1}$ and
$f_1^N(t)=f_{N}^{1}(t)=\left(-i\sin t\right)^{N-1}$. With these expressions at
hand, we can now predict the sequence of QST fidelity for an arbitrary readout
timing sequence $\left\{t_n,t_{n-1},\dots,t_2,t_1\right\}$ using
\cref{eq_fid_gen} and its simplified forms valid for free fermion models as
given, e.g.\ in \cref{eq_Favg2,eq_Favg3}. As an illustrative example, let us
consider the case where the readout time is affected by a constant shift
$\delta$ from the ideal PST time $\tau$ for all $t_i$. 

\begin{figure}
    \centering
    \includegraphics[width=\linewidth]{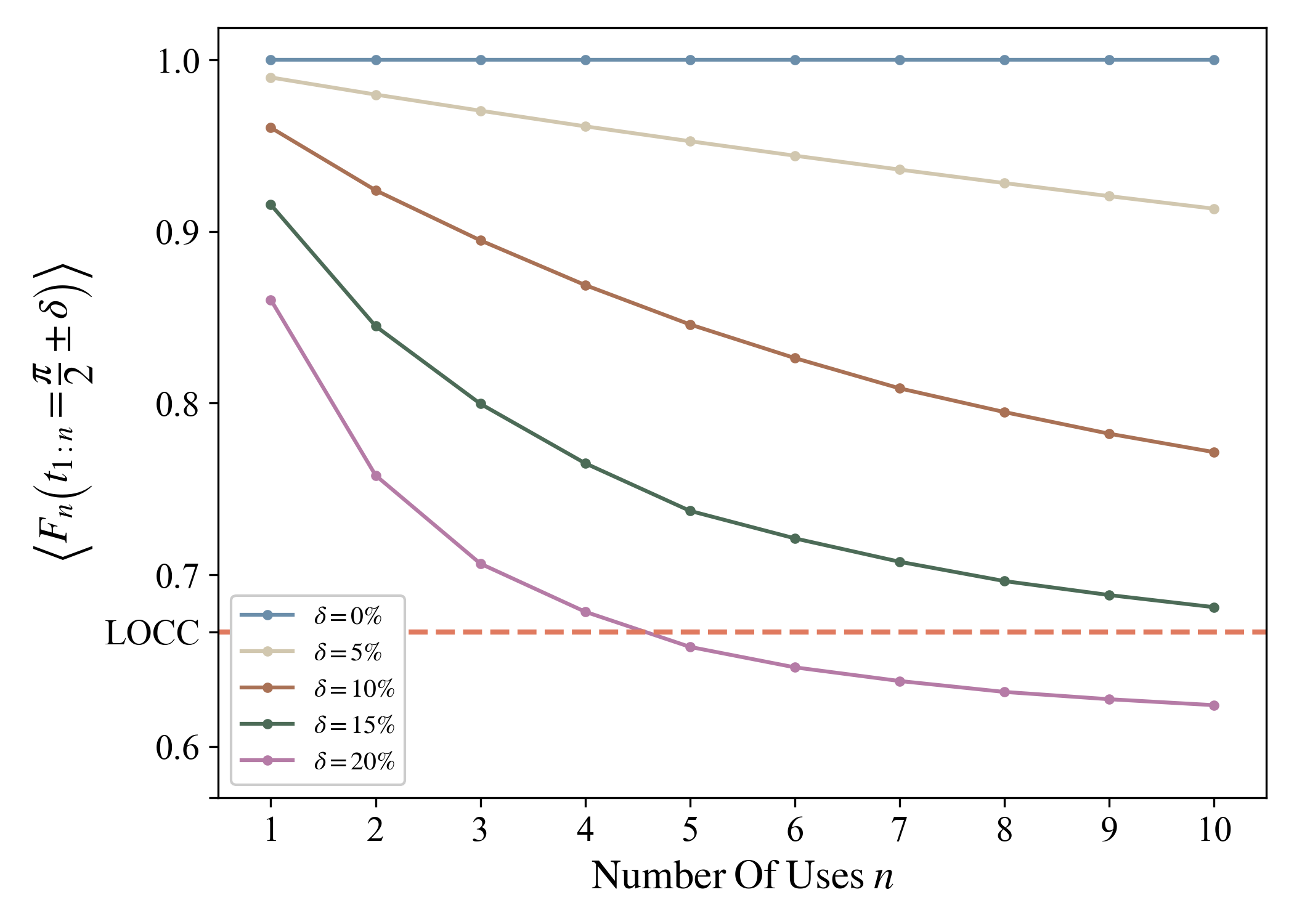}
    \caption{Average $n\nthscript{th}$-use fidelity for a chain of fixed length $N=6$ in
        \cref{eq_Chris} for different readout timing errors $\delta$. The red dashed
        line reports the LOCC limit $\frac{2}{3}$. Apart from the ideal scenario of
        $\delta=0\%$, each subsequent use of the quantum channel lowers the
        fidelity.}
    \label{fig:scaling_with_uses}
\end{figure}

In Fig.~\ref{fig:scaling_with_uses} we see that already for $\delta=5\%$, the average fidelity reduces to
about 0.91 after 10 uses.
Next we consider the $n\nthscript{th}$-use
fidelity scaling with the length of the quantum channel. 
\begin{figure}
    \centering
    \includegraphics[width=1\linewidth]{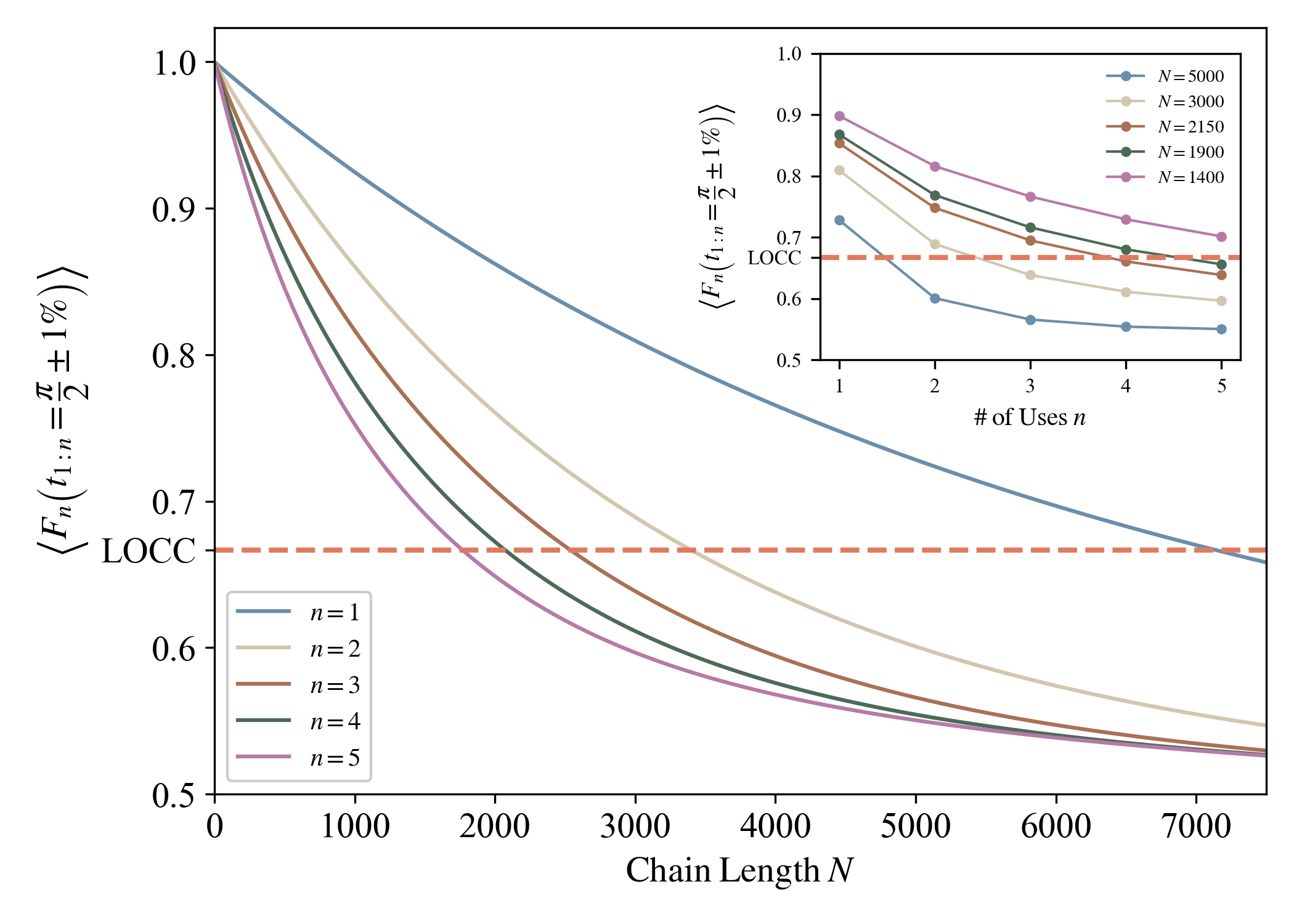}
    \caption{Average fidelity for a fixed readout timing error of $\delta=1\%$
        for different chain lengths in \cref{eq_Chris}. Each curve represents a
        different number of uses $n=1,2,\dots,5$. Already after a few uses, chains
        in the order of $10^3$ sites fall below the LOCC limit (red dashed line).
        The inset shows the average $n\nthscript{th}$-use fidelity along vertical lines of
        the main plot for selected lengths $N$.}
    \label{fig:scaling_with_length}
\end{figure}
In Fig.~\ref{fig:scaling_with_length}, we report the value of the average
fidelity up to the \nth{5}-use for chain lengths up to $N=7500$ for a readout error
of $\delta=1\%$. We see that longer chains are more sensitive to readout timing
errors, e.g.\ for a chain of $N=2150$, already the $\nth{4}$-use yields a
fidelity that is attainable by Local Operations and Classical
    Communications (LOCC) \cite{Horodecki2009Review,chitambar2014}.

Finally, we consider the case where the quantum channel is used to distribute entanglement. The formalism developed for QST can be readily applied for entanglement distribution exploiting the quantum dynamical map formalism~\cite{Society2004,Lorenzo2022}: $\hat{\rho}^{\left(s'r\right)}_{ij}=\Phi_{ij}^{nm}(t)\hat{\rho}^{\left(s's\right)}_{nm}$, with $i,j,n,m=0,1,2,3$. The map, acting on the qubits $\left(s's\right)$ and producing the output of the qubits $\left(s'r\right)$, is obtained by  $\Phi^{\left(s's\right)}(t)=\mathds{1}^{\left(s'\right)}\otimes \Lambda^{\left(s\right)}$, with $\Lambda^{\left(s\right)}$ denoting the single-qubit map acting on the sender. Whereas the first use of the quantum channel $\Phi_1$ results in the standard amplitude damping channel~\cite{Bose01012007},  we directly derive
the following expression for the \nth{2}-use dynamical map as the product
of two completely positive trace-preserving (CPTP) maps:
\begin{equation}
\Phi_{2}=\Phi_{\mathrm{GAD}}\left(\gamma_{2},p_{2}\right)\Phi_{\mathrm{PD}}\left(\lambda_{2}\right)\label{eq:decom}
\end{equation}
where $\Phi_{\mathrm{GAD}}$ and $\Phi_{\mathrm{PD}}$ are the superoperators
corresponding respectively to a generalized amplitude damping 
and a dephasing channel \cite{Khatri2020,Devetak2005,Puch2021}. 

In the computational basis they read, 
\begin{equation}
\Phi_{\mathrm{GAD}}\!\left(\gamma_{2},p_{2}\right)=\!\!\left(\begin{array}{cccc}
1\!-\!\frac{1}{2}\!\!\left|B_{2}\right|^{2} & 0 & 0 & 1\!-\!\frac{1}{2}\!\!\left|B_{2}\right|^{2}\!\!\!-\!\left|\!f_{1}^{N}\!\!\left(t_{2}\right)\!\right|^{2}\\
0 & f_{1}^{N}\!\!\left(t_{2}\right)^{\ast} & 0 & 0\\
0 & 0 & f_{1}^{N}\!\!\left(t_{2}\right) & 0\\
\frac{1}{2}\!\!\left|B_{2}\right|^{2} & 0 & 0 & \frac{1}{2}\!\!\left|B_{2}\right|^{2}\!\!\!+\!\left|\!f_{1}^{N}\!\!\left(t_{2}\right)\!\right|^{2}
\end{array}\right)
\end{equation}
and 
\begin{equation}
\Phi_{\mathrm{PD}}\left(\lambda_{2}\right)=\left(\begin{array}{cccc}
1 & 0 & 0 & 0\\
0 & A_{1}(t_1) & 0 & 0\\
0 & 0 & A_{1}(t_1) & 0\\
0 & 0 & 0 & 1
\end{array}\right)
\end{equation}
with damping parameter $\gamma_{2}=1-\left|f_{1}^{N}\left(t_{2}\right)\right|^{2}$, mixing probability
$p_{2}=\frac{1}{\gamma_{2}}[1-\frac{1}{2}\left|B_{2}\right|^{2}-\left|f_{1}^{N}\left(t_{2}\right)\right|^{2}]$, and dephasing 
$\lambda_{2}=1-\left(A_{1}(t_1)\right)^{2}$.
 
\begin{figure}
        \centering
        \includegraphics[width=1\linewidth]{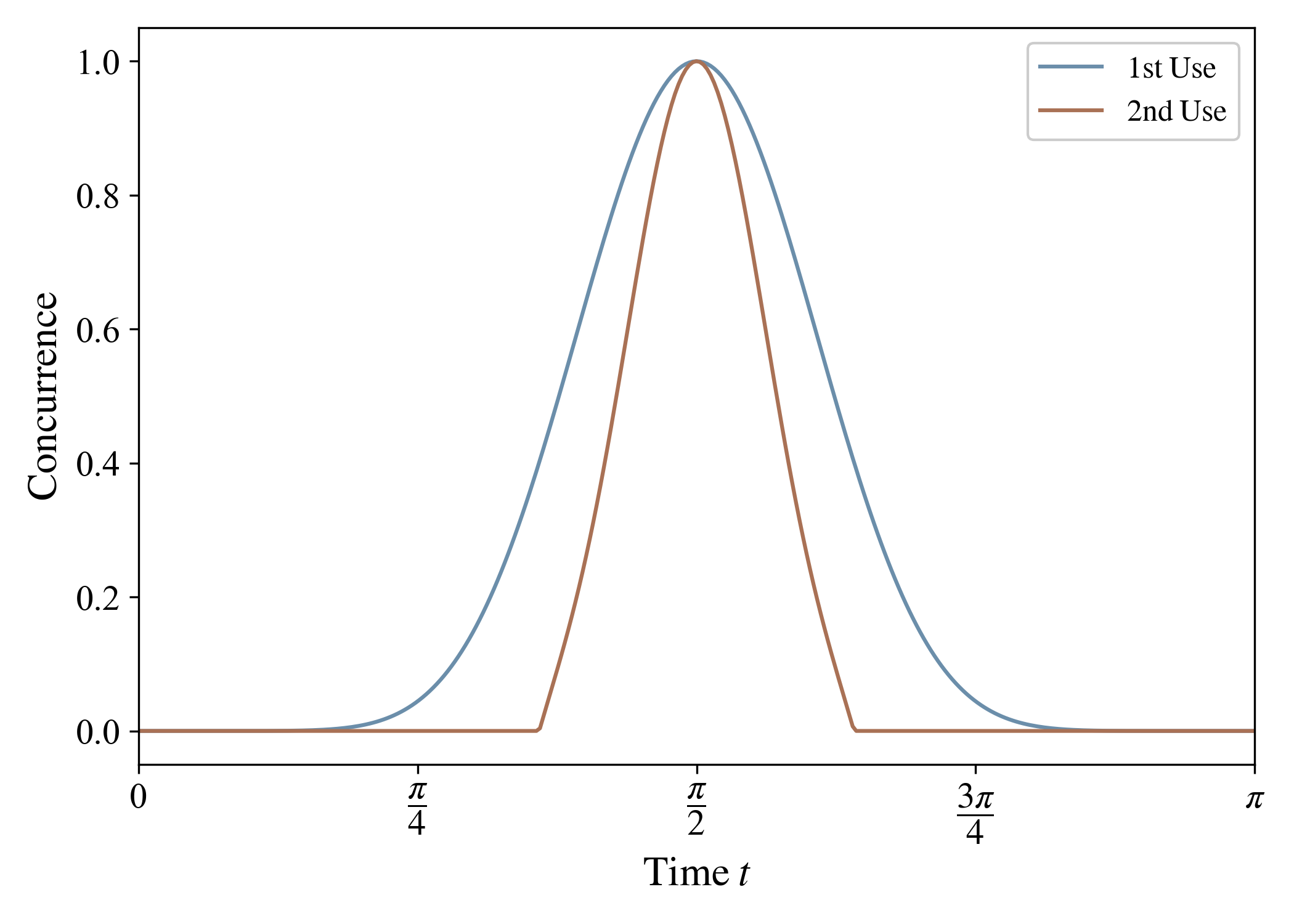}
        \caption{Entanglement distribution for the \nth{1}- (blue curve) and
            \nth{2}-use (red curve) across the quantum channel modeled by
            \cref{eq_Chris} with $N=10$ as a function of time. Whereas the
            \nth{1}-use admits a non-zero concurrence between $s'$ and $r$ at any
            $t\neq 0$ or $\pi$, the \nth{2}-use admits only a reduced time-window
            for entanglement distribution.}
        \label{fig:concurrence}
    \end{figure}

Employing the bottleneck inequality \cite{Wolf2007} over \cref{eq:decom}, and then exploiting the  convexity of the quantum channel capacity $Q\left(\cdot\right)$ for degradable channels  \cite{Wolf2007,Khatri2020}, the following bound for the quantum capacity of the \nth{2}-use channel is straightforward:
\begin{equation}
Q\left(\Phi_2\right)\leq p_{2}I_{c}\!\left[\Phi_{\mathrm{AD}}\!\left(\gamma_{2},0\right)\right]\!+\!\left(\!1\!-\!p_{2}\!\right)I_{c}\!\left[\Phi_{\mathrm{AD}}\!\left(\gamma_{2},1\right)\right] \label{eq:fundbounds}
\end{equation} for any $\gamma_{2}< 1/2$ and where the channel coherent information $I_{c}\left[\Phi\right]=\max_{\rho_s\in\mathcal{B}\left(\mathcal{H}_s\right)}\left\{I_{c}\left(\rho_s,\Phi\right)\right\} $
stands for the maximum over input states of the standard coherent information 
\cite{Barnum1998,NielsenChuang2010}. For channel uses at equal time intervals $t_{2}=t_{1}$, since $Q\left(\Phi_1\right)= I_{c}\left[\Phi_{\mathrm{AD}}\!\left(\gamma_{2},1\right)\right]=I_{c}\left[\Phi_{\mathrm{AD}}\!\left(\gamma_{2},0\right)\right] $ \cite{Khatri2020}, the upper bound in \cref{eq:fundbounds} guarantees that $Q\left(\Phi_2\right)\leq Q\left(\Phi_1\right)$
for any $\gamma_{2}=\gamma_{1}=1-\left|f_{1}^{N}\left(t_{1}\right)\right|^{2} < 1/2$ and from the bottleneck inequality we have $Q\left(\Phi_2\right)=Q\left(\Phi_{\mathrm{AD}}\!\left(\gamma_{2},p\right)\right)=0$ for $\gamma_{2}=\gamma_{1} \geq 1/2$ due to antidegradability of $\Phi_{\mathrm{AD}}$ \cite{Khatri2020}. Hence, the second use of
the channel reduces its ability to transmit quantum information
for any protocol, establishing a limitation on the efficiency of
entanglement distribution through  repeated uses in uniform intervals.
We illustrate the above by implementing the protocol of
    Ref.~\cite{Bose2003}, where entanglement initially shared by the qubit pair
    $\left(s',s\right)$ is transferred to $\left(s',r\right)$. As shown in
    Fig.~\ref{fig:concurrence}, the first use yields finite concurrence
    ~\cite{woottersEntanglementFormationArbitrary1998} for all times $t\neq 0$
    or $\pi$, while when $t_2=t_1$ the second use exhibits broad
    time windows of vanishing entanglement transfer, exposing the destructive role of
    channel memory.
 
\section{Conclusions}\label{sec_concl}
We have investigated an analytic framework
for evaluating memory effects on the QST average fidelity and on the distributed
entanglement due to repeated uses of a $U(1)$-symmetric quantum channel. We
obtained an exact expression of the average fidelity for an arbitrary number of
uses of the quantum channel in terms of single-particle transition
amplitudes involving only the sender and receiver sites when the quantum channel
is represented by a quadratic Hamiltonian in its fermionic representation. We
showed that the average fidelity becomes increasingly sensitive to readout
timing errors with the number of uses and, interestingly, commonly investigated
entanglement distribution protocols, e.g.\ based on the PST quantum channel, may
be inefficient already after the first use in the presence of unavoidable
readout timing errors. Our results are readily applicable to a variety of
quantum state and entanglement distribution
protocols~\cite{Kay2006,apollaro99FidelityBallistic2012,ronkeAndersonLocalisationSpin2016a,
    zhouQuantumStateTransfer2024, juniorQuantumStateTransfer2025} and may have an
important role in discriminating which one is more robust in terms of
multiple uses, contributing thus both to the scalability and the energy
efficiency of quantum technologies where QST and entanglement distribution are
key primitives.

\section{Acknowledgements} 
HZ and TJGA are grateful to Karol \.{Z}yczkowski for useful discussions. The authors acknowledge funding from Grant Agreement No.~DTP-2024-13, Project ATTESTER financed by the Xjenza Malta, and the Malta Digital Innovation Authority, through the `R\&I Thematic Programmes: Digital Technologies Programme'. HZ also acknowledges funding by the Tertiary Education Scholarships Scheme (TESS).
\vspace{0.5cm}
\section{Appendix}
In the case of a $U(1)$-symmetric  Hamiltonian that is not quadratic in the fermionic representation, the $n^{\mathrm{th}}$-use average fidelity is given by

\begin{widetext}
    \begin{align}\label{eq.general-avg}
        \average{F_{n}(t_n;t_1)} & =\frac{1}{2}+ \frac{1}{3}\left| \left[\sum_{k=0}^{\mathrm{min}\{n-1,N\}} \sum_{p_{[k]},q_{[k]}} \sum_{p'_{[k]}} \rho^{(n-1)}_{p_{[k]},q_{[k]}}(t_{n-1};t_1) f_{1 p_{[k]}}^{p'_{[k]} N}(t_{n}){f_{q_{[k]}}^{p'_{[k]}}(t_{n})}^{*} \right]\right|                                                                                        \\
                                        & +\frac{1}{6}\left[ \sum_{k=0}^{\mathrm{min}\{n-1,N\}} \sum_{p_{[k]},q_{[k]}} \rho^{(n-1)}_{p_{[k]},q_{[k]}}(t_{n-1};t_1)\left(\sum_{p'_{[k]}} f_{p_{[k]}}^{p'_{[k]}}(t_{n}) {f_{q_{[k]}}^{p'_{[k]}}(t_{n})}^{*} - \sum_{p'_{[k+1]}} f_{1 p_{[k]}}^{p'_{[k+1]}}(t_{n}){f_{1 q_{[k]}}^{p'_{[k+1]}}(t_{n})}^{*} \right)  \right]\nonumber
    \end{align}
\end{widetext}

where $p_{[k]}$ and $q_{[k]}$ refer to length k sets of indices over the chain
 excluding the sender and receiver sites, whereas sets $p'_{[k]}$ also
include the sender site. The $\rho^{(n-1)}_{p_{[k]},q_{[k]}}$ refer to the
density channel elements of the $k\nthscript{th}$ particle sector of the previous use.
Finally, the summation over $k$ runs up to $n-1$ while the chain is unsaturated,
after which $n-1$ is replaced with its maximum value $N$.

\bibliography{n_use.bib}

\end{document}